\documentclass[prb,aps,twocolumn,showpacs,superscriptaddress,floatfix]{revtex4}
\usepackage[cp1251]{inputenc}
\usepackage[english]{babel}
\usepackage{epsfig}
\usepackage{epstopdf}
\usepackage{amsmath}
\usepackage{amssymb}
\usepackage{amsfonts}
\usepackage{mathptmx}
\usepackage{eucal}
\usepackage{bm}

\graphicspath{{Pictures1/}}

\begin{document}

\title{Dynamic properties of asymmetric double Josephson junction stack with
quasiparticle imbalance}
\author{S.~V.~Bakurskiy}
\affiliation{Skobeltsyn Institute of Nuclear Physics, Lomonosov Moscow State University
1(2), Leninskie gory, Moscow 119234, Russian Federation}
\affiliation{Moscow Institute of Physics and Technology, Dolgoprudny, Moscow Region,
141700, Russian Federation}
\affiliation{MIREA - Russian Technological University, 119454 Moscow, Russian Federation}
\author{A.~A.~Neilo}
\affiliation{Faculty of Physics, Lomonosov Moscow State University, 119992 Leninskie
Gory, Moscow, Russian Federation}
\author{N.~V.~Klenov}
\affiliation{Faculty of Physics, Lomonosov Moscow State University, 119992 Leninskie
Gory, Moscow, Russian Federation}
\affiliation{Skobeltsyn Institute of Nuclear Physics, Lomonosov Moscow State University
1(2), Leninskie gory, Moscow 119234, Russian Federation}
\affiliation{Moscow Institute of Physics and Technology, Dolgoprudny, Moscow Region,
141700, Russian Federation}
\affiliation{MIREA - Russian Technological University, 119454 Moscow, Russian Federation}
\affiliation{All-Russian Research Institute of Automatics n.a. N.L. Dukhov (VNIIA),
127055, Moscow, Russian Federation}
\author{I.~I.~Soloviev}
\affiliation{Skobeltsyn Institute of Nuclear Physics, Lomonosov Moscow State University
1(2), Leninskie gory, Moscow 119234, Russian Federation}
\affiliation{Moscow Institute of Physics and Technology, Dolgoprudny, Moscow Region,
141700, Russian Federation}
\affiliation{MIREA - Russian Technological University, 119454 Moscow, Russian Federation}
\author{M.~Yu.~Kupriyanov}
\affiliation{Skobeltsyn Institute of Nuclear Physics, Lomonosov Moscow State University
1(2), Leninskie gory, Moscow 119234, Russian Federation}
\affiliation{Moscow Institute of Physics and Technology, Dolgoprudny, Moscow Region,
141700, Russian Federation}
\date{\today }

\begin{abstract}
We study analytically and numerically the influence of the quasiparticle
charge imbalance on the dynamics of the asymmetric Josephson stack formed by
two inequivalent junctions: the fast capacitive junction $JJ_{1}$ and slow
non-capacitive junction $JJ_{2}$. We find, that the switching of the fast
junction into resistive state leads to significant increase of the effective
critical current of the slow junction. At the same time, the initial
switching of the slow junction may either increase or decrease the effective
critical current of the fast junction, depending on ratio of their
resistances and the value of the capacitance. Finally, we have found that
the slow quasiparticle relaxation (in comparison with Josephson times) leads
to appearance of the additional hysteresis on current-voltage
characteristics.
\end{abstract}

\maketitle

%Features of the quasiparticle accumulation in the systems of
%unequivalent Josephson junctions }

\section{Introduction}

Josephson junctions with multilayer structures in a weak link region between
superconducting (S) electrodes are of considerable interest for rapidly
developing superconducting spintronics \cite{Eschrig1, Linder1, Blamire1,
Soloviev1}. The important class of these devices contains thin
superconducting layers s inside this area. These spacers additionally
support superconducting correlations inside the weak link and permits to
increase a critical current of the junctions compare to that with normal
spacers \cite{Baek, Gingrich}. For instance, SFsFS spin valves \cite%
{AlidoustSFSFS, KrasnovSFSFS, OuassouSFSFS, Klenov} has in a weak link FsF
three-layer or periodic FsF structure formed by different ferromagnets (F).
The critical current of such devices depends on the mutual orientation of
adjacent F layers magnetization vectors. The next class of devices is based
on the SIsFS structures \cite{Larkin, Bakurskiy2013, Ruppelt}. Their weak
place contains an insulator (I) and only one F layer. The SIsFS spin valves
can combine the properties of a fast and energy-efficient element of logic
circuits SIs with the possibility of long-term information storage in the
form of the direction of the magnetization vector of the F-layer \cite%
{Vernik, BakurskiyAPL} or in the unconventional phase states of the middle
s-layer \cite{Bakurskiy2016, Bakurskiy2017, Bakurskiy2018}. The other types
of layers also can be considered. The ferromagnetic insulators (FI) \cite%
{Blamire2017, Giazotto2018} or multilayers insulator-ferromagnetic metal F-I 
\cite{Pugach2011} can be used to obtain magnetic properties without strong
suppression in s-layer due to inverse proximity effect. Implementation of
topological insulators (TI) \cite{Brinkman2018} may add into the system $%
4\pi $ periodic component of the current-phase relation.

The practical applications of such devices meet a number of difficulties
associated with the lack of understanding of the dynamic processes occurring
in them. The accurate consideration of this problem requires the solution of
the unequilibrium equations of the microscopic theory of superconductivity 
\cite{WBZ}. It is a very difficult task even in symmetric structures that do
not contain a superconductor in the weak link region \cite{Brinkman2003}.

In this paper, we analyze the dynamic processes within a simpler
phenomenological approach. In it, there are two lumped Josephson junctions
connecting in series via thin intermediate s layer. This s layer is
spatially homogeneous and its thickness, $d_{S},$ is of order of coherence
length, $\xi _{S},$ and much smaller than the London penetration depth, $%
\lambda $. The critical currents, $I_{C1,C2},$ normal resistances, $R_{1,2},$
and capacitances, $C_{1,2},$ of the junctions are different and junction's
dynamics is described by modified resistive shunt model (MRSJ ) taking into
account coupling processes between the junctions.

\begin{figure}[t]
\begin{minipage}[h]{0.99\linewidth}
\center{\includegraphics[width=0.99\linewidth] {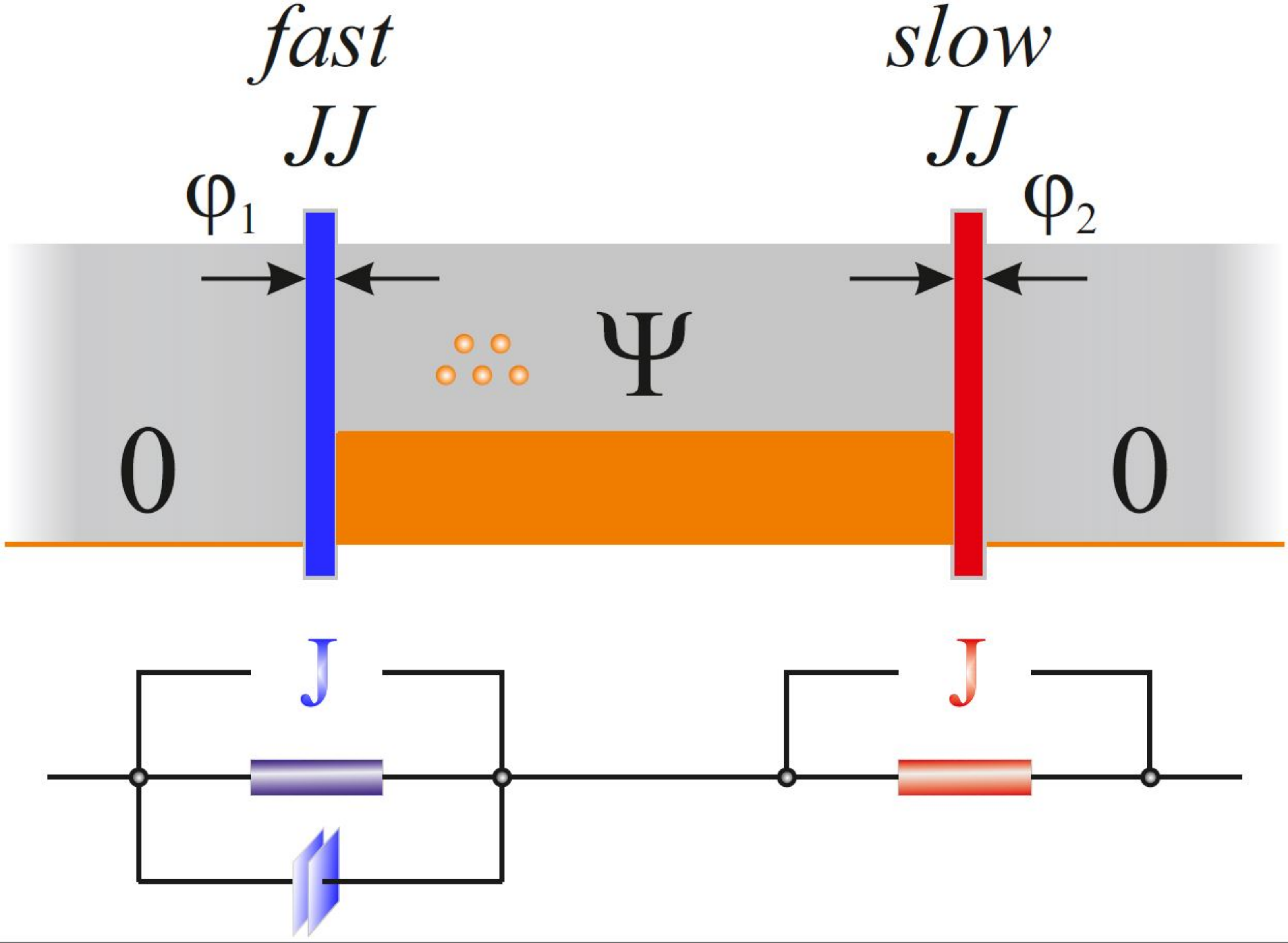}}
\end{minipage}
\caption{Sketch of the asymmetric double junction stack with equivalent
scheme of the circuit in the frame of RSJ-model.}
\label{Scetch}
\end{figure}

In carrying out the necessary modifications of the RSJ model, we used
extensive material obtained earlier in the analysis of processes in the
stacks of identical tunnel Josephson junctions and multilayer
high-temperature superconducting materials \cite{Nevirkovets,
Shafranjuk1996, Sakai1993, Ustinov, Kleiner2, Kleiner, Bulaevskii1996, Hu,
Koyama1996, Matsumoto1999, Machida1999, Shukrinov2006, Shukrinov2007,
Shukrinov2012, Ryndyk1997, Artemenko1997, Ryndyk1998, Shafranjuk1999,
Ryndyk1999, Ryndyk2003, Ryndyk2005, Volkov2007}. In these studies, three
mechanisms of coupling between Josephson contacts in the stack were
identified. They are inductive interaction between adjacent junctions \cite%
{Nevirkovets, Shafranjuk1996, Sakai1993, Ustinov, Kleiner2, Kleiner,
Bulaevskii1996, Hu}, a charge accumulation of condensate \cite{Koyama1996,
Matsumoto1999, Machida1999, Shukrinov2006, Shukrinov2007, Shukrinov2012} and
a quasiparticle accumulation \cite{Ryndyk1997, Artemenko1997, Ryndyk1998,
Shafranjuk1999, Ryndyk1999, Ryndyk2003, Ryndyk2005, Volkov2007}.

The first two are not relevant for our study. The inductive interaction is
important if $d_{S}>\lambda $ and the width of the stack is larger than
Josephson penetration depth $W>\lambda _{J},$ while the charge accumulation
of condensate occurs when the intermediate s-layer is thinner than the Debye
charge screening length $\lambda _{D}$. These conditions are not met in our
model.

The quasiparticle accumulation in the intermediate s layer may occur if at
least one of the junction in stack is in a resistive state. Under this
condition the full current sets in the s film should contain both normal and
superconducting components. If the thickness $d_{S}$ of the s layer is
substantially less than the length of the energy relaxation of the
quasiparticles injected into it, then a charge imbalance arises in the s
film due to the different population of the electron and hole branches of
the energy spectrum. The total charge quasi-neutrality is achieved at the
same time due to superconducting electrons. It leads to the difference of
the gradient-invariant potential of the condensate $\Phi =\Psi $ in the s
film from its value in the bulk S electrodes, which is supposed to be in
equilibrium, that is having electropotential $\Phi =0.$

\section{Model}

Following Ryndyk \cite{Ryndyk1997, Ryndyk1998} we may write the system of
equations of MRSJ model in the form 
\begin{eqnarray}
\frac{\partial \varphi _{1}}{\partial t} &=&V_{1}+\Psi ,  \label{fi1} \\
i &=&\sin \varphi _{1}+V_{1}+\beta \frac{\partial V_{1}}{\partial t},
\label{Jeqi1} \\
\frac{\partial \varphi _{2}}{\partial t} &=&V_{2}-\Psi ,  \label{fi2} \\
i &=&a\sin \varphi _{2}+rV_{2},  \label{Jeq4} \\
\tau _{Q}\frac{\partial \Psi _{1}}{\partial t} &=&-\Psi +\tilde{\kappa}%
\left( I_{1}^{qp}-I_{2}^{qp}\right) =-\Psi +\kappa \left(
V_{1}-rV_{2}\right) ,  \label{KinEq0} \\
\kappa &=&\frac{\tau _{Q}}{2e^{2}R_{1}N_{0}d_{s}}.  \label{kap0}
\end{eqnarray}%
Here times $t$ and $\tau _{Q}$, currents $i$, $I_{1}^{qp},I_{2}^{qp},$
voltages $V_{1},$ $V_{2}$ and potential $\Psi $ are normalised on $\omega
_{c1}^{-1}$, critical current $I_{C1}$, and characteristic voltage, $%
I_{C1}R_{1},$ respectively,$\,\ $ $\tau _{Q}$ is time of quasiparticle
relaxation, $\kappa $ is coupling parameter, $e$ - electron charge, $N_{0}$
- density of states of the s film, $I_{1}^{qp}$ and $I_{2}^{qp}$ are
quasiparticle currents across the junctions. We also introduce the notations 
$\beta =C_{1}2\pi I_{C1}R_{1}^{2}/\Phi _{0},$ $r=R_{1}/R_{2},$ $%
a=I_{C2}/I_{C1}$ and assume that capacitance of the second junction is
negligibly small and can be omitted. Below we additionally restrict ourself
by considering the most interesting for us case in which $i$ is independent
in time bias current and there is a large difference between junction's
normal resistances, $r\gg 1,$ while their critical currents have the same
order of magnitude. Then the characteristic frequency of the first junction $%
\omega _{c1}=2\pi I_{C1}R_{1}/\Phi _{0}$ is much larger than that of the
second one. In this sense we call the first junction as "fast" (implying as
it is regular tunnel SIs junction), and call the second junction as "slow"
(it can be more complicated structure). The figure \ref{Scetch} shows a
schematic representation of the structure under study.

\section{Fast quasiparticle relaxation, $\tau _{Q}\ll 1$}

In the limit of fast quasiparticle relaxation in the intermediate s layer in
comparison with the characteristic Josephson times $\tau _{Q}\ll 1$ we can
neglect the left side in the kinetic equation (\ref{KinEq0}) and rewrite (%
\ref{fi1}), (\ref{fi2}) in the form 
\begin{eqnarray}
\frac{\partial \varphi _{1}}{\partial t} &=&V_{1}+\kappa \left(
V_{1}-rV_{2}\right) ,  \label{bEq1} \\
\frac{\partial \varphi _{2}}{\partial t} &=&V_{2}-\kappa \left(
V_{1}-rV_{2}\right) .  \label{bEq4}
\end{eqnarray}%
Equations (\ref{bEq1}), (\ref{bEq4}) mean that the interaction between the
fast and slow junction is reduced to the redistribution of the electric
potential difference between them%
\begin{eqnarray}
V_{1} &=&q\frac{\partial \varphi _{1}}{\partial t}+rp\frac{\partial \varphi
_{2}}{\partial t},  \label{V1} \\
V_{2} &=&m\frac{\partial \varphi _{2}}{\partial t}+p\frac{\partial \varphi
_{1}}{\partial t},  \label{V2}
\end{eqnarray}%
where 
\begin{equation}
p=\frac{\kappa }{1+\kappa +\kappa r},\quad q=\frac{1+\kappa r}{1+\kappa
+\kappa r},\quad m=\frac{1+\kappa }{1+\kappa +\kappa r}.
\end{equation}

Making use of (\ref{bEq1}), (\ref{bEq4}) we can rewrite (\ref{Jeqi1}), (\ref%
{Jeq4}) in the closed for $\varphi _{1}$ and $\varphi _{2}$ forms 
\begin{eqnarray}
i &=&\sin \varphi _{1}+q\frac{\partial \varphi _{1}}{\partial t}+rp\frac{%
\partial \varphi _{2}}{\partial t}+  \label{EqFi1} \\
&&+\beta q\frac{\partial ^{2}\varphi _{1}}{\partial t^{2}}+\beta rp\frac{%
\partial ^{2}\varphi _{2}}{\partial t^{2}},  \notag
\end{eqnarray}%
\begin{equation}
i=a\sin \varphi _{2}+mr\frac{\partial \varphi _{2}}{\partial t}+pr\frac{%
\partial \varphi _{1}}{\partial t}.  \label{EqFi2}
\end{equation}

\subsection{Slow junction in the superconducting state, $a>1$ \label{LargeA}}

Consider the situation when the fast junction is in the resistive state,
while the slow one is in the superconducting state and suppose additionally
that $\beta \gg 1.$ 
%We try to find the solution of them in the limit of large capacitance of the
%fast junction $\beta >>1$ and in the region of parameters, when the fast
%junction is in the resistive state, while the slow junction is in the
%superconducting state. At this stage we assume, that the critical current of
%the slow junction is larger than fast's one $a>1$.

Under these conditions we may find solution of equations (\ref{EqFi1}), (\ref%
{EqFi2}) in the form 
\begin{equation}
\varphi _{1}=\Omega _{1}t+\tilde{\varphi}_{1};~\varphi _{2}=\varphi _{20}+%
\tilde{\varphi}_{2},  \label{SmallFi}
\end{equation}%
where $~\tilde{\varphi}_{1},\tilde{\varphi}_{2}\propto \beta ^{-1}\ll 1$ -
are small periodic in time functions, while $\Omega _{1}~$and $\varphi _{20}$
are independent on time frequency of Josephson oscillations of the fast
junction and phase difference across the slow junction. Substitution of the (%
\ref{SmallFi}) into (\ref{EqFi1}) leads to

\begin{eqnarray}
i &=&\sin \Omega _{1}t+\tilde{\varphi}_{1}\cos \Omega _{1}t+q\Omega _{1}+q%
\frac{\partial \tilde{\varphi}_{1}}{\partial t}+  \label{EqFi1C} \\
&&+rp\frac{\partial \tilde{\varphi}_{2}}{\partial t}+\beta q\frac{\partial
^{2}\tilde{\varphi}_{1}}{\partial t^{2}}+\beta rp\frac{\partial ^{2}\tilde{%
\varphi}_{2}}{\partial t^{2}}.  \notag
\end{eqnarray}%
After averaging over the period oscillation in equation (\ref{EqFi1C}), we
arrive at 
\begin{equation}
i=\left\langle \tilde{\varphi}_{1}\cos \Omega _{1}t\right\rangle +q\Omega
_{1}  \label{iav}
\end{equation}%
and in the zero approximation on $\beta ^{-1}$ for $\Omega _{1}$ we get. 
\begin{equation}
\Omega _{1}\simeq iq^{-1}.  \label{Vsr}
\end{equation}%
Taking (\ref{Vsr}) into account, in the next approximation from (\ref{EqFi1C}%
) we have%
\begin{equation}
\frac{\partial ^{2}(q\tilde{\varphi}_{1}+pr\tilde{\varphi}_{2})}{\partial
t^{2}}=-\frac{\sin \Omega _{1}t}{\beta }  \label{Deq_comb}
\end{equation}%
resulting in%
\begin{equation}
q\tilde{\varphi}_{1}+pr\tilde{\varphi}_{2}=\frac{\sin \Omega _{1}t}{\beta
\Omega _{1}^{2}},  \label{Fi1res}
\end{equation}%
\begin{equation}
q\frac{\partial \tilde{\varphi}_{1}}{\partial t}+pr\frac{\partial \tilde{%
\varphi}_{2}}{\partial t}=\frac{\cos \Omega _{1}t}{\beta \Omega _{1}},
\label{dFi1res}
\end{equation}

Substitution of (\ref{Fi1res}), (\ref{dFi1res}) into the equation (\ref%
{EqFi2}) leads to%
\begin{eqnarray}
i &=&a\sin \left( \varphi _{20}+\tilde{\varphi}_{2}\right) +rm\frac{\partial 
\tilde{\varphi}_{2}}{\partial t}+\frac{rp}{q}i+ \\
&&+\frac{rp}{q}\frac{\cos \Omega _{1}t}{\beta \Omega _{1}}-\frac{r^{2}p^{2}}{%
q}\frac{\partial \tilde{\varphi}_{2}}{\partial t},  \notag
\end{eqnarray}%
which transforms after some algebra into%
\begin{eqnarray}
\frac{i}{(1+\kappa r)} &=&\frac{rp}{q}\frac{\cos \Omega _{1}t}{\beta \Omega
_{1}}+a\sin \left( \varphi _{20}\right) +  \label{Fi2sl} \\
&&+a\tilde{\varphi}_{2}\cos \left( \varphi _{20}\right) +\frac{r}{(1+\kappa
r)}\frac{\partial \tilde{\varphi}_{2}}{\partial t}.  \notag
\end{eqnarray}%
Averaging over the period oscillation in equation (\ref{Fi2sl}) gives the
magnitude of effective critical current $i_{C2}^{\ast }$ of the slow junction%
\begin{equation}
i=i_{C2}^{\ast }\sin \left( \varphi _{20}\right) ,\quad i_{C2}^{\ast
}=a(1+\kappa r),  \label{IC_eff}
\end{equation}%
which exceeds the intrinsic value of the critical current state $a.$

Taking into account (\ref{IC_eff}) from (\ref{Fi2sl}) we further get%
\begin{equation}
-pr\frac{\cos \Omega _{1}t}{\beta i}=a\cos \left( \varphi _{20}\right) 
\tilde{\varphi}_{2}+\frac{r}{\left( 1+\kappa r\right) }\frac{\partial \tilde{%
\varphi}_{2}}{\partial t},  \label{Fi2var}
\end{equation}%
where for the bias current $i$ set in the positive direction $(i>0)$ 
\begin{equation}
\cos \left( \varphi _{20}\right) =\frac{\sqrt{a^{2}\left( 1+\kappa r\right)
^{2}-i^{2}}}{a\left( 1+\kappa r\right) }.
\end{equation}%
The solution of (\ref{Fi2var}) is%
\begin{equation}
\tilde{\varphi}_{2}(t)=-\frac{\kappa r}{\beta \Omega _{1}}\cos \left( \Omega
_{1}t-\varphi _{20}\right) ,  \label{Fi2Res}
\end{equation}%
leading to 
\begin{equation}
\tilde{\varphi}_{1}(t)=\frac{\sin \Omega _{1}t}{q\beta \Omega _{1}^{2}}+%
\frac{\kappa pr^{2}}{\beta i}\cos \left( \Omega _{1}t-\varphi _{20}\right) .
\label{Fi1Res}
\end{equation}%
Substitution of (\ref{Fi1Res}) into (\ref{iav}) gives the correction to
frequency of oscillations $\Omega _{1}$ in the next approximation in $\beta
^{-1}$%
\begin{equation}
\Omega _{1}=\frac{i}{q}-\frac{\kappa pr^{2}}{2q\beta i}\cos \left( \varphi
_{20}\right) .  \label{Ombetter}
\end{equation}

\begin{figure}[t]
\begin{minipage}[h]{0.99\linewidth}
\center{\includegraphics[width=0.99\linewidth]{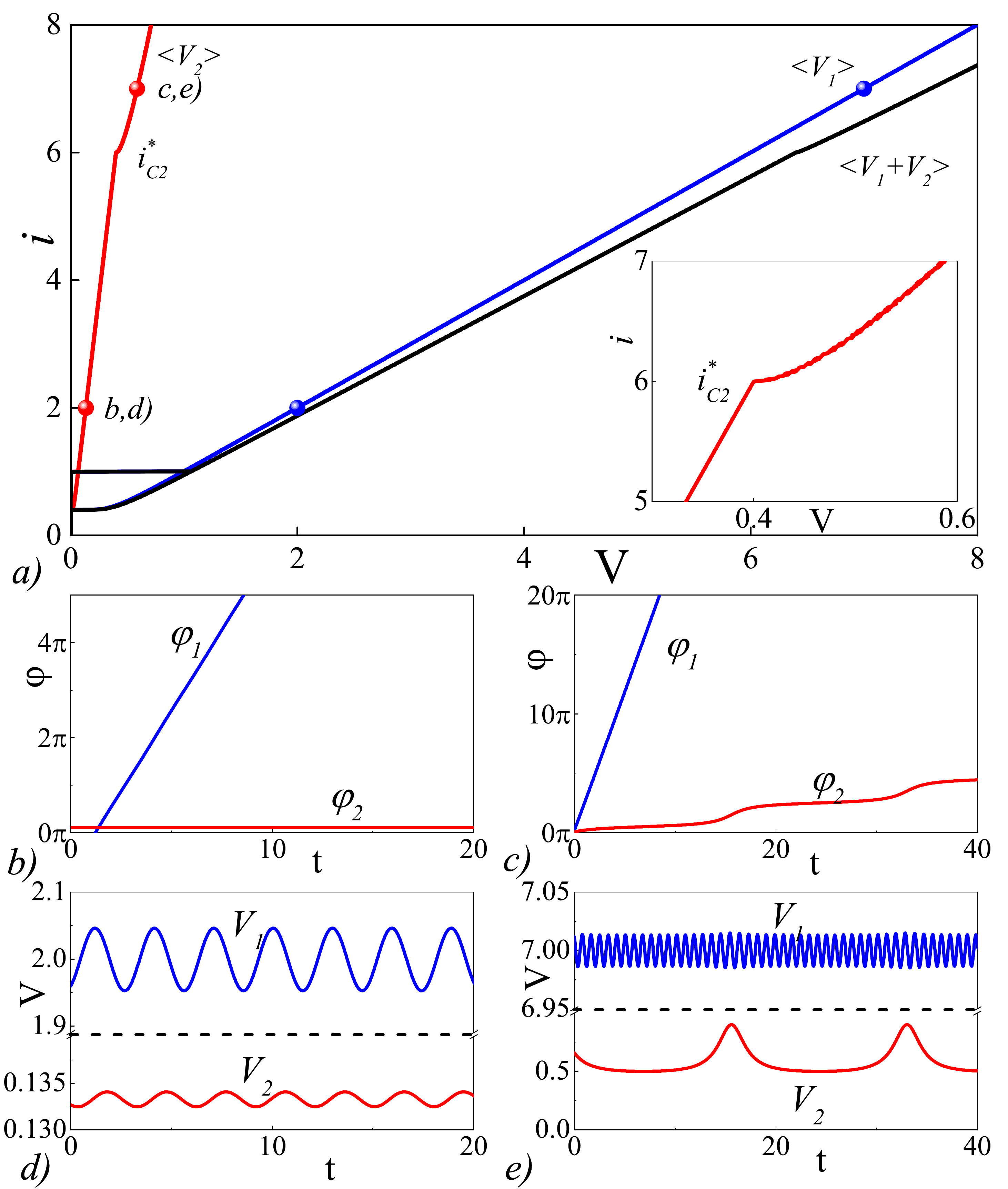}}
\end{minipage}
\caption{I-V characteristic of the asymmetric Josephson stack with coupling $%
\protect\kappa =0.2$. The black line demonstrates the IV dependence of the
whole system, the blue line corresponds fast junction and red line describes
slow junction. The panels b-e) show phase and voltage dynamics calculated
b,c) at the current $i=2$ below the switch of the slow junction into
resistive state and d, e) at the current $i=7$ over that switch. The other
parameters are $r=10$, $\protect\beta =10$, and $a=2$.}
\label{IV1}
\end{figure}

As a result, the bias current\ on the fast junction can be represented as
the sum of independent on time normal, $(q\Omega _{1}),$ and superconducting
parts 
\begin{equation}
i=q\Omega _{1}+\frac{\kappa pr^{2}}{2q\beta i}\cos \left( \varphi
_{20}\right) .\ 
\end{equation}%
The normal current components of the bias current is not fully converted
into the superconducting one inside the s film, so that there is an
accumulation of quasiparticles inside the film. As a consequence, a voltage
drop 
\begin{eqnarray}
V_{2} &=&m\left( \Omega _{1}+\frac{\cos \Omega _{1}t}{q\beta \Omega _{1}}-%
\frac{\kappa pr^{2}}{q\beta }\sin \left( \Omega _{1}t-\varphi _{20}\right)
\right) \\
&&+p\frac{\kappa r}{\beta }\sin \left( \Omega _{1}t-\varphi _{20}\right) 
\notag
\end{eqnarray}%
occurs on a slow junction, despite the fact that the total current $i$ is
less than its critical one. It means that the slow junction is biased by the
superposition of the superconducting and normal current components. As soon
as the normal current does not affect the critical one, while the sum of
these independent on time components must be equal to external bias $i,$ the
critical state of the slow junction must be achieved at larger magnitude of $%
i=i_{C2}^{\ast }.$ The critical current enhancement, $i_{C2}^{\ast
}-a=a\kappa r$\ is exactly equal to independent in time normal component of
bias current across the slow junction.

To generalize these properties for the case of finite $\beta $, we
numerically solved Eqs.(\ref{bEq1})-(\ref{bEq4}) for $\beta =10$. \ The
calculations have been done for coupling parameter $\kappa =0.2$, the ratio
of resistances $r=10$, and critical current ratio $a=2$. The Fig. \ref{IV1}a
shows current - voltage characteristics (IVC) of the considered stucture,
where black line respects to whole structure, while blue and red lines
correspond to fast $\left\langle V_{1}\right\rangle \ $and slow $%
\left\langle V_{2}\right\rangle $\ junctions respectively. The points on the
curves mark the positions on the IVC at $i=2$ and $i=7$ for which the time
dependences of the voltages $V_{1,2}$ and phase differences $\varphi _{1,2}$
across the contacts are shown in the Fig. \ref{IV1}b-Fig. \ref{IV1}e.

At the point marked by the letter $b$ the slow junction is in the
superconducting state. As it is seen in Fig. \ref{IV1}b phase difference $%
\varphi _{2}$ undergoes oscillations with the frequency $\Omega _{1}$ around
constant over time value, while $\varphi _{1}$ grows linearly with time. The
voltage drops $V_{1,2}$ are also oscillated with the frequency $\Omega _{1}$
relatively the appropriate constant over time values, as it is seen from
Fig.\ref{IV1}d. At $i=7$ both junctions are in resistive state. Figures \ref{IV1}c,e give the time evolutions of $\varphi _{1,2}$ and $V_{1,2}$
at $i=7$ respectively.

The numerical results confirm the analytical estimates. In the full
accordance with (\ref{V1})-(\ref{V2}) when the bias current $i$ exceeds the
unity (the critical current of fast junction), there is voltage drop across
the structure and it\ is redistributed between fast and slow junctions.\ It
is also seen that the slow junction starts to generate at $i=i_{C2}^{\ast
}=a\left( 1+\kappa r\right) =6.$ Below this point, the voltage of the slow
junction \emph{\ }has oscillating component with a frequency of the fast
junction, while over the critical current $i_{C2}^{\ast }$ it has much
smaller frequency.

\subsection{Fast junction in the superconducting state, $a<1$ \label{SmallA}}

\begin{figure}[t]
\begin{minipage}[h]{0.99\linewidth}
\center{\includegraphics[width=0.99\linewidth]{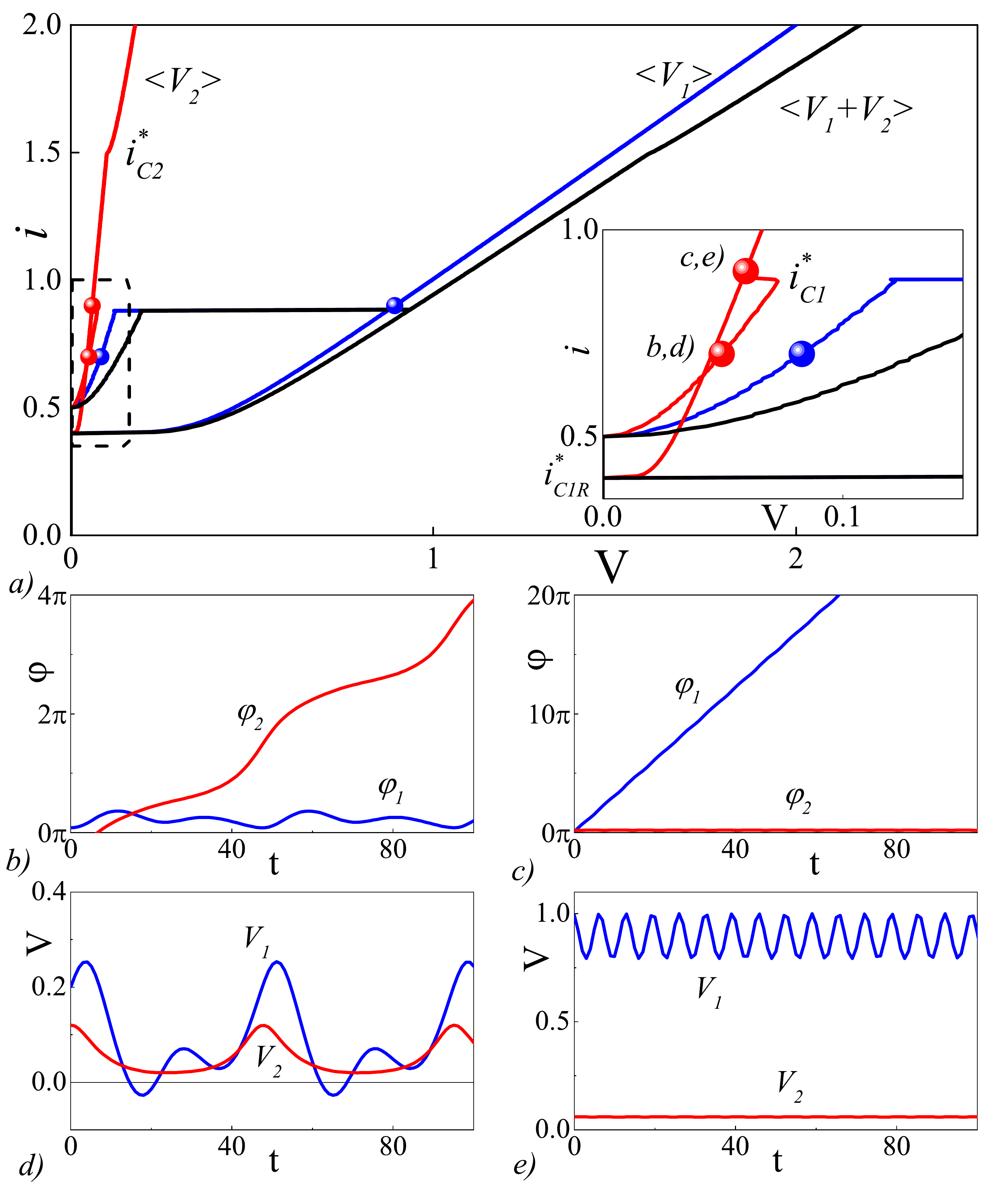}}
\end{minipage}
\caption{a) I-V characteristic of the asymmetric Josephson stack with
coupling $\protect\kappa =0.2$. The black line demonstrates the IV
dependence of the whole system, the blue line corresponds fast junction and
red line describes slow junction. The panels b-e) show phase and voltage
dynamics calculated b,c) at the current $i=0.7$ below the switch of the fast
junction into resistive state and d, e) at the current $i=0.9$ over that
switch. The other parameters are $r=10$, $\protect\beta =10$, and $a=0.5$.}
\label{IV2}
\end{figure}

In this case, application of the bias current to the structure leads to the
switching of the slow junction into the resistive state. At the same time,
the fast junction is in a superconducting state despite the fact that in
accordance with (\ref{V1}) it has a voltage drop overlaid by the flow of a
normal current component across it. In the limit $\beta \Omega _{2}\gg 1,$
where, $\Omega _{2},$ is the frequency of Josephson oscillations of the slow
junction. From (\ref{EqFi1}) it follows that as the first approximation on $%
(\beta \Omega _{2})^{-1}$ we can assume that 
\begin{equation}
q\frac{\partial ^{2}\varphi _{1}}{\partial t^{2}}=-pr\frac{\partial
^{2}\varphi _{2}}{\partial t^{2}}  \label{BetaD}
\end{equation}%
and after integration obtain%
\begin{equation}
\frac{\partial \varphi _{1}}{\partial t}=\frac{pr}{q}\left( \Omega _{2}-%
\frac{\partial \varphi _{2}}{\partial t}\right) .  \label{Fi12c}
\end{equation}%
The integration constant in (\ref{Fi12c}) has been determined from the
condition for the absence of intrinsic Josephson generation in fast
junction. Substitution of (\ref{Fi12c}) into (\ref{EqFi1}) leads to the
equation containing $\varphi _{2}$ only 
\begin{equation}
i-\eta \Omega _{2}=a\sin \varphi _{2}+\frac{r}{\left( 1+\kappa r\right) }%
\frac{\partial \varphi _{2}}{\partial t},\quad \eta =\frac{p^{2}r^{2}}{q}.
\label{Fi2lowJ}
\end{equation}%
Solution of this equation \cite{Aslamazov1968} has the form%
\begin{equation}
\frac{d\varphi _{2}}{dt}=\frac{u\left( 1+\kappa r\right) }{r}\left[
1+2\sum_{n=1}^{\infty }\left( \frac{a}{i-\eta \Omega _{2}+ua}\right)
^{n}\cos \frac{ua\left( 1+\kappa r\right) }{r}nt\right] ,  \label{dfi2dt}
\end{equation}%
where%
\begin{equation}
u=\sqrt{(i-\eta \Omega _{2})^{2}/a^{2}-1}  \label{u}
\end{equation}%
is the average voltage across the slow junction. Carrying out in (\ref%
{dfi2dt}) averaging over the oscillation period for $\Omega _{2}$, we have

\begin{equation}
\Omega _{2}=\frac{i^{2}-a^{2}}{i\eta +a\sqrt{\eta ^{2}+r^{2}\left(
i^{2}-a^{2}\right) /\left( 1+\kappa r\right) ^{2}}}.  \label{omegaLow}
\end{equation}%
Expressions (\ref{Fi12c}), (\ref{u}) and (\ref{omegaLow}) determine the time
evolution of a phase difference $\varphi _{1}$ on the fast junction

\begin{equation}
\varphi _{1}=\varphi _{10}-\frac{2\kappa r}{\left( 1+\kappa r\right) a}%
\sum_{n=1}^{\infty }\frac{1}{n}\left( \frac{a}{i-\eta \Omega _{2}+ua}\right)
^{n}\sin \frac{ua\left( 1+\kappa r\right) }{r}nt  \label{fi1fast}
\end{equation}%
where $\varphi _{10}$ is independent on time $t$ phase difference across the
fast junction. Averaging in (\ref{EqFi1}) over period of slow junction
frequency oscillations gives 
\begin{equation}
i=\left\langle \sin \varphi _{1}\right\rangle +pr\Omega _{2}.
\label{fi1eqav}
\end{equation}%
From (\ref{fi1fast}), (\ref{fi1eqav}) it follows that the critical current
of the fast junction can be achieved at $i_{C1}^{\ast }<1.$ Indeed, even in
the case when we restrict ourselves only to the first term of the series
with respect to $n$ we get that 
\begin{equation}
i=\sin \varphi _{10}J_{0}\left( \frac{2\kappa r}{\left( 1+\kappa r\right)
\left( i-\eta \Omega _{2}+ua\right) }\right) +pr\Omega _{2},  \label{Bes_est}
\end{equation}%
where $J_{0}(z)\leq 1$ is the zero order Bessel function.

The critical current $i_{C1}^{\ast }$ is determined from (\ref{Bes_est}) at $%
\sin \varphi _{10}=1$ and it is affected by two physical mechanisms. The
first one relates to the term $pr\Omega _{2}$ and correspond to appearance
of the normal component of current through the fast junction similarly with
Sec.\ref{LargeA}. It tends to increase the $i_{C1}^{\ast }$ up to $(1+\kappa
)$. The second impact related with coefficient $J_{0}(z)\leq 1$ tends to
decrease the critical current and it is explained by the presence of the
oscillations of the phase $\varphi _{1}$, which have significant amplitude
unlike the previous subsection.

Figure \ref{IV2} \ shows the results of numerical calculations follow from
Eqs.(\ref{bEq1})-(\ref{bEq4}) for the set of parameter relevant to the
considered limit, namely, $\kappa $ $=0.2,$ $r=10$, $\beta $ $=10$ and $%
a=0.5 $. Black line in Fig. \ref{IV2}a is the IVC of the whole structure.
Blue and red curves are IVC of the fast and the slow junctions,
respectively. As shown in the Fig. \ref{IV2}a inset gives in more detail the
initial part of IVC located in the dotted rectangle. The points on the
curves mark the positions on the IVC at $i=0.7$ and $i=0.9$ for which the
time dependences of the voltages $V_{1,2}$ and phase differences $\varphi
_{1,2}$ across the contacts are shown in the Fig. \ref{IV2}b-Fig. \ref{IV2}e.

\begin{figure}[t]
\begin{minipage}[h]{0.99\linewidth}
\center{\includegraphics[width=0.99\linewidth]{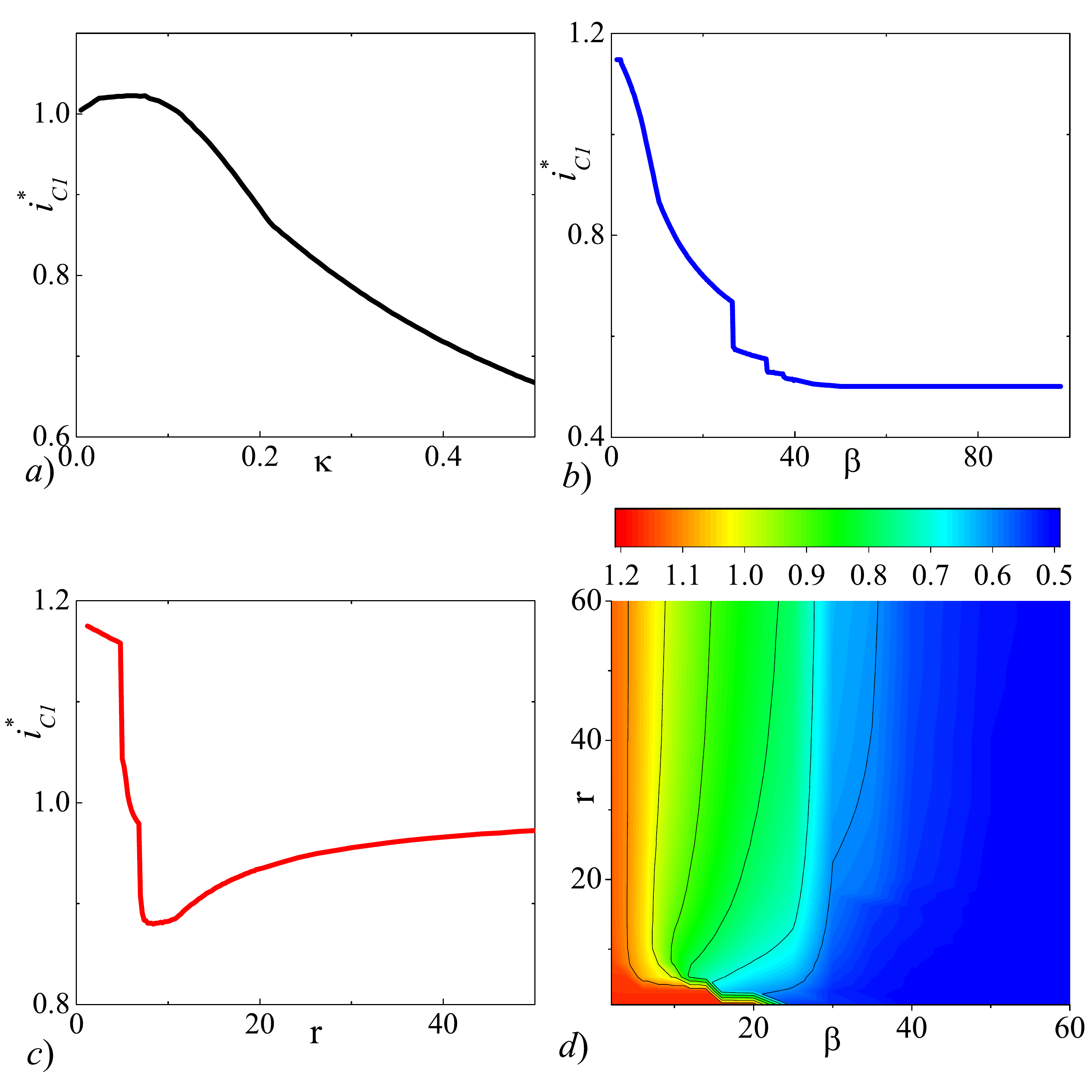}}
\end{minipage}
\caption{ Dependence of the effective critical current $i_{C1}^{\ast }$ of
the fast junction versus a) coupling parameter $\protect\kappa$, b)
parameter $\protect\beta$ and c) ratio of resistances $r$. Panel
d) demonstrates $i_{C1}^{\ast }$ on the phase plain in coordinates $\protect%
\beta$ and $r$. The other parameters are $\protect\kappa=0.2$, $r=10$, $%
\protect\beta =10$ and $a=0.5$.}
\label{Param}
\end{figure}

It can be seen from the Fig. \ref{IV2}a that as soon as the bias current $i$
exceeds the critical current of slow junction $a,$ a voltage drop occurs on
both contacts. It increases with the\ $i$ growth if $i\leq i_{C1}^{\ast
}\approx 0.89.$ Typical evolutions of $\varphi _{1,2}$ and $V_{1,2}$ at $%
i=0.7$ is demonstrated at Fig. \ref{IV2}b and Fig. \ref{IV2}c, respectively.
It is seen that in the considered bias current interval $a\leq i\leq
i_{C1}^{\ast }$ there are the time oscillations of phase difference $\varphi
_{2}$ across the slow junction superimposed on its linear growth, while the
phase difference $\varphi _{1}$ oscillates with respect to a time-constant
value. At $i=i_{C1}^{\ast }\approx 0.89$ there is a transition of the fast
junction into resistive state, which, due to the large value of parameter $%
\beta $, is accompanied by a jump on the IVC to a region of high voltages.
This circumstance substantially changes the balance of quasiparticle
currents flowing into the s layer. If, at $i\leq i_{C1}^{\ast }$, the
quasiparticles were injected into the s layer through slow contact, then at $%
i>i_{C1}^{\ast }$, a substantially large number of quasiparticles from the
fast transition enters this layer and there is a change sign of potential $%
\Psi .$ This results in increase of slow junction critical curent to the
value $i_{C2}^{\ast }=a(1+\kappa r),$ that is up to $i_{C2}^{\ast }=3a$ for
the chosen values of $\kappa $ $=0.2\ $and $r=10.$ \ The slow junctions goes
into superconducting state with independent in time $\varphi _{2}$ and $%
V_{2} $ (see Fig. \ref{IV2}d,e, which are provided the results of
calculations for $i=0.9>i_{C1}^{\ast }).$ From the Fig. \ref{IV2}c,e it is
also easy to see that at $i=0.9$ the phase difference $\varphi _{1}$
increases linearly with time, and the voltage drop $V_{1}$ oscillates around
a constant value. At $i=i_{C2}^{\ast }=3a=1.5$ the slow junction contact
switches to a resistive state, it is evident from the kink in its IVC.
During the reverse motion along the I -- V characteristic in the direction
of decreasing the bias current $i$, the slow contact is first transitioned
to the superconducting state at $i=i_{C2}^{\ast }=a(1+\kappa r),$ while the
fast junction makes a similar transition abruptly at a current $%
i=i_{C1R}^{\ast }\approx 0.4<a$.

Interestingly, for large $\beta $ and $\kappa $, the effective critical
current $i_{C1}^{\ast }$ can become less than the critical current of the
slow junction $a$. In this case transition of the slow junction into
resistive state initiate the transition to the same state of the fast
junction, the process takes place during the time $t\sim \Omega _{2}^{-1}$.
The last transition switches the slow junction into the superconducting
state and for $t\gtrsim \Omega _{2}^{-1}$ only fast junction is in the
resistive state. Exactly this regime is predicted analytically by (\ref%
{Bes_est}) in the case $\beta \rightarrow \infty $ for parameters $\kappa $ $%
=0.2,$ $r=10$ and $a=0.5$. Fig.\ref{Param} permits to check it, showing
dependencies versus $\kappa $, $\beta $ and$\ r$ on the panels a), b) and c)
respectively. While any of those parameters is small, that the critical
current is larger then unity $i_{C1}^{\ast }>1$ and tends to value $%
(1+\kappa \dot{)}$. Increase of the parameters leads to the decrease of the $%
i_{C1}^{\ast }$, with significant drops on $i_{C1}^{\ast }(\beta )$ and $%
i_{C1}^{\ast }(r)$ dependencies. These drops are related with fullfilment of
the condition $\beta \Omega _{2}\gg 1$ and lead to the qualitative change of
the phase dynamic. At very large $\beta \,>50$\ the critical current $%
i_{C1}^{\ast }$ reaches the value $a=0.5$ and becomes permanent. At Fig. \ref%
{Param}d we show the $i_{C1}^{\ast }$ dependence on the $r$-$\beta $ plane
and demonstrate that the latter regime appears at large $\beta $ for a wide
range of $r$.

\section{{Large relaxation time $t_{Q}>>1$}}

\begin{figure}[t]
\begin{minipage}[h]{0.99\linewidth}
\center{\includegraphics[width=0.99\linewidth]{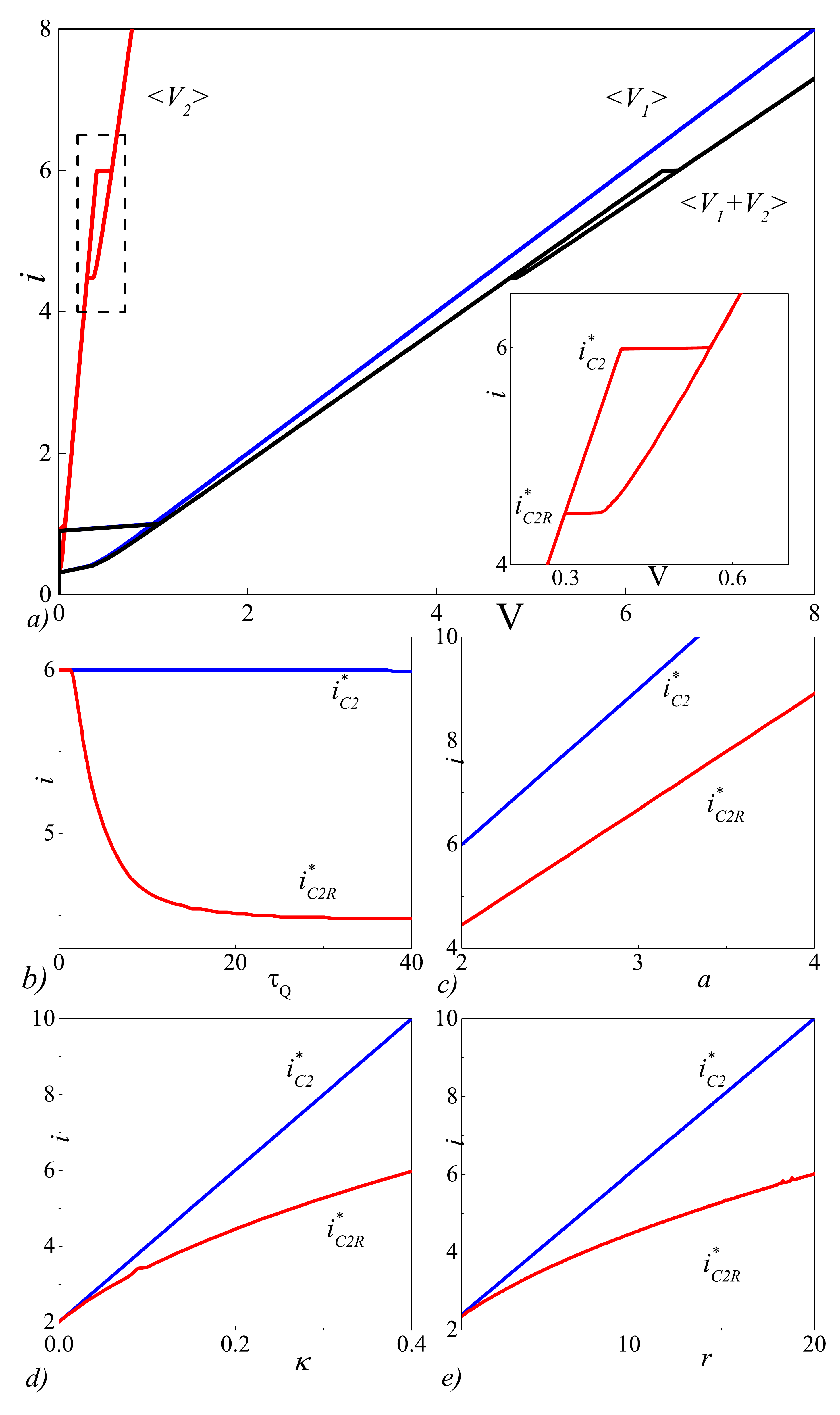}}
\end{minipage}
\caption{a) I-V characteristic of the asymmetric Josephson stack for slow relaxation $\protect\tau_{Q} =50$. The black line
demonstrates the IV dependence of the whole system, the blue line
corresponds fast junction and red line describes slow junction. Inset
enlarges the area around critical current of the slow junction. b-e)
Critical currents $i_{C2}^{\ast }$ and $i_{C2R}^{\ast }$ b) as function of
relaxation time $\protect\tau_{Q}$, c) critical current $a$, d) coupling
parameter $\protect\kappa$ and e) resistivity ratio $r$. The other
parameters was taken from the set $\protect\tau_{Q}=50$, $r=10$, $\protect%
\kappa=0.2$, $\protect\beta =10$ and $a=2$. }
\label{IV_TQ}
\end{figure}

In this approximation, the potential $\Psi $ in the s layer does not have
time to react to the instantaneous change in voltage at the junctions and is
determined by their time-averaged values 
\begin{equation}
\Psi =\kappa \left( \left\langle V_{1}\right\rangle -r\left\langle
V_{2}\right\rangle \right) .  \label{Psi1TQL}
\end{equation}%
The correction to this solution of the equation (\ref{KinEq0}) has the order
of $\tau _{Q}^{-1}$ and is proportional to the difference of oscillating in
time components $V_{1}-\left\langle V_{1}\right\rangle $ and $%
V_{2}-\left\langle V_{2}\right\rangle $ of voltage drops across the
contacts. Substitution of (\ref{Psi1TQL}) into (\ref{fi1}), (\ref{fi2}) gives%
\begin{eqnarray}
\frac{\partial \varphi _{1}}{\partial t} &=&V_{1}+\kappa \left( \left\langle
V_{1}\right\rangle -r\left\langle V_{2}\right\rangle \right) ,  \label{fi1sr}
\\
\frac{\partial \varphi _{2}}{\partial t} &=&V_{2}-\kappa \left( \left\langle
V_{1}\right\rangle -r\left\langle V_{2}\right\rangle \right) .  \label{fi2sr}
\end{eqnarray}

To demonstrate the specific features of the behavior of the structure under
study in the limit of large $\tau _{Q}$, it is enough to consider the case $%
a>1.$ At $1<i<a$ the slow junction is in the superconducting state, while
the fast one has switched to the quasiparticle branch of the I -- V
characteristic to the region of high voltages, where $i\approx \left\langle
V_{1}\right\rangle $ and 
\begin{equation}
\varphi _{1}=\Omega _{1}t+\tilde{\varphi}_{1}.
\end{equation}%
The voltages $V_{1}$ and $V_{2}$ are almost permanent with a small periodic
correction. In this way $V_{1,2}\approx \left\langle V_{1,2}\right\rangle $
and behaviour of the system is similar to that discussed in Subsec.\ref%
{LargeA}. In particular, the critical current of the switching of the slow
junction into resistive state, $i_{C2}^{\ast }=a(1+\kappa r),$ is exactly
the same as it was found in Subsec.\ref{LargeA}.

However, after transition of the slow junction into the resistive regime
this similarity is broken. In this case, at $i$ slightly larger $%
i_{C2}^{\ast }$ the slow junction generates periodic components $\tilde{%
\varphi}_{1,2}$,%
\begin{equation}
\varphi _{1}=\Omega _{1}t+\tilde{\varphi}_{1};~\varphi _{2}=\Omega _{2}t+%
\tilde{\varphi}_{2},
\end{equation}%
which have an order of unity. This provides the significant difference
between instantaneous values of $V_{1,2}\ $and averaged $\left\langle
V_{1,2}\right\rangle $ voltage. In this case, the averages are coupled
similarly with (\ref{V1})-(\ref{V2}) of Subsec \ref{LargeA}. 
\begin{equation}
\left\langle V_{1}\right\rangle =q\Omega _{1}+rp\Omega _{2}
\end{equation}%
\begin{equation}
\left\langle V_{2}\right\rangle =m\Omega _{2}+p\Omega _{1}
\end{equation}%
while the equations for periodic component are similar with equations for
separate junctions with modified effective bias currents%
\begin{equation}
i+p\left( \Omega _{1}-r\Omega _{2}\right) =\sin \left( \varphi _{1}\right) +%
\frac{\partial \varphi _{1}}{\partial t}+\beta \frac{\partial ^{2}\varphi
_{1}}{\partial t^{2}}  \label{Ieq_tq}
\end{equation}%
\begin{equation}
i-rp\left( \Omega _{1}-r\Omega _{2}\right) =a\sin \left( \varphi _{2}\right)
+r\frac{\partial \varphi _{2}}{\partial t}  \label{Ieq_tq_2}
\end{equation}

Since the fast junction stays on the resistive branch of IVC, we can neglect
averaged part of $\sin \left( \varphi _{1}\right) $ term in Eq. \ref{Ieq_tq}
and get the equality 
\begin{equation}
i=q\Omega _{1}+pr\Omega _{2},  \label{tq_ass1}
\end{equation}%
which transforms the Eq. (\ref{Ieq_tq_2}) into%
\begin{equation}
\frac{i+\kappa r^{2}\Omega _{2}}{\left( 1+\kappa r\right) }=a\sin \left(
\varphi _{2}\right) +r\frac{\partial \varphi _{2}}{\partial t}
\label{tq_ass2}
\end{equation}%
having solution \cite{Aslamazov1968}%
\begin{equation}
\frac{d\varphi _{2}}{dt}=\frac{u}{r}\left[ 1+2\sum_{k>0}\left( \frac{a}{%
i_{eff2}+ua}\right) ^{k}\cos \frac{ua\left( 1+\kappa r\right) }{r}kt\right] ,
\label{Sol_tq1}
\end{equation}%
\begin{equation}
u=\sqrt{i_{eff2}{}^{2}/a^{2}-1};~i_{eff2}=\frac{i+\kappa r^{2}\Omega _{2}}{%
\left( 1+\kappa r\right) }.
\end{equation}%
After time averaging in (\ref{Sol_tq1}) we get the equation for $\Omega _{2}$%
\begin{equation}
\Omega _{2}=\frac{u}{r}=\frac{1}{r}\sqrt{\left( \frac{i+\kappa r^{2}\Omega
_{2}}{a\left( 1+\kappa r\right) }{}\right) ^{2}-1},
\end{equation}%
which has the solution 
\begin{equation}
\Omega _{2}=\frac{i\kappa r+i_{C2}^{\ast }\sqrt{i^{2}+\kappa
^{2}r^{2}-i_{C2}^{\ast 2}}}{r\left( i_{C2}^{\ast 2}-\kappa ^{2}r^{2}\right) }
\label{tq_mode2}
\end{equation}%
The slow junction stays in the resistive state until the expression under
the root crosses zero. Then, the slow junction returns into the
superconducting state at bias current $i=i_{C2R}^{\ast }$, 
\begin{equation}
i_{C2R}^{\ast }=\sqrt{i_{C2}^{\ast 2}-\kappa ^{2}r^{2}}.  \label{tq_crit_2}
\end{equation}

Numerical solution of the (\ref{fi1})-(\ref{KinEq0}) for finite values of
parameters qualitatively confirms the analytical estimates. The I-V curve of
the considered system for the large relaxation time $t_{Q}=50$ is
demonstrated in the Fig. \ref{IV_TQ}a (the other parameters are the same
with Fig. \ref{IV1}: $a=2,$ $r=10$, $\kappa =0.2,$ $\beta =10$). Inset of
Fig. \ref{IV_TQ}a enlarges the vicinity of the critical point for the slow
junction $V_{2}$. It is clear, that its transition to the resistive state
occurs abruptly when the bias current reaches the value $i_{C2}^{\ast
}=a(1+\kappa r)=6.$ However, during the decrease of the bias current, the
slow junction stays in the resistive state until the current $i_{C2R}^{\ast
}\approx 4.4$. In Fig. \ref{IV_TQ}b we demonstrate the evolution of the
critical $i_{C2}^{\ast }$ and return current $i_{C2R}^{\ast }$ as a function
of $t_{Q}$. The return current starts to decrease when the $t_{Q}$ is
comparable with $\omega _{C1}^{-1}=1$, and reaches the asymptote when $t_{Q}$
significantly exceeds the $\omega _{C2}^{-1}=r=10$. The dependencies of the $%
i_{C2}^{\ast }$ and $i_{C2R}^{\ast }$ on parameters $a,$ $\kappa $ and $r$
are shown in the Fig.\ref{IV_TQ}c-e. The $i_{C2R}^{\ast }$ curves have the
shape close to that followed from (\ref{tq_crit_2}) with linear dependence
versus $a$, and root-like versus $\kappa $ and $r$. The exact values of the
return current is smaller than analytical estimates, due to limited validity
of approximation (\ref{tq_ass1}) at the finite $\beta $, and, thus, the
hysteresis loop becomes more noticeable.

\section{Discussion}

In the paper we consider analytically and numerically the dynamics of the
asymmetric Josephson stack with two inequivalent junctions: the fast
capacitive junction $JJ_{1}$ and\ slow non-capacitive junction $JJ_{2}$. The
quasiparticle imbalance in the thin superconducting layer between junctions
leads to significant changes of the system dynamical properties:

1) If the fast junction is in the resistive state, and slow junction is in
the superconducting state, then the effective critical current $i_{C2}^{\ast
}$ of the slow junction is growing up. This effect is stronger for junctions
with higher ratio of resistances.

2) In the case of slow junction in resistive state and fast junction in
superconducting state, the effective critical current $i_{C1}^{\ast }$ of
the fast junction may be either increased or decreased depending on
parameters of the system. Numerical solution demonstrates that its effective
critical current is increased for the weak coupling $\kappa $, small
resistance ratio $r$ and small parameter $\beta \,$, while at the large
parameters it is decreased.

3) If the quasiparticle relaxation is slower than Josephson times $t_{Q}\gg
\omega _{C1,2}^{-1}$, the coupling is leading to hysteresis on the
current-voltage characteristic of slow non-capacitive junction. The
quasiparticle injection through the slow junction leads to increase of its
generation frequency $\Omega _{2}$ and provides some kind of resistive
branch of \ IVC for non-capacitive junction.

Features on the IVC at subgap voltages similar to those obtained in this
study were previously observed in double-barrier SI$_{1}$sI$_{2}$S structures 
\cite{Balashov, Tolpygo, Nevirkovets2}. However, they were not the subject
of study in these structures. It is for this reason; a quantitative
comparison of the predictions of the developed model with these experimental
data is difficult. For instance, it is unclear how to distinguish the
modified critical currents of the junctions $i_{C1,2}^{\ast }$ from
their truly critical currents $i_{C1,2}$. However, it may be possible if one
of the junctions has widely variable parameters, for instance, as in
Josephson spin-valve devices. One can smoothly modify their critical current
with remagnitization of the ferromagnetic layer, providing the transition
between the regimes of Sec.\ref{LargeA} and \ref{SmallA}. It gives a
possibility to measure as well as the truly critical current as the modified
one for the both junctions.

Even more intriguing case occurs for the junction with controllable $0$-$\pi 
$ transition \cite{Ryazanov1, Ryazanov2006}, at which the critical current
of the junction changes on the orders of magnitude. Moreover, the hysteretic
nature of considered effect can lead to the different dynamical states
inside $0$-$\pi $ transition performed with or without bias current.

\textbf{Acknowledgments.} The authors acknowledge helpful discussions with
Yu. M. Shukrinov, V. V. Bol'ginov and D. A. Ryndyk. The analytical study was
supported by RFBR (18-32-00672 mol-a) and numerical calculations were done
with support of Russian Science Foundation (17-12-01079).

\end{document}